# Microresonator soliton frequency combs via cascaded Brillouin scattering

Hao Zhang[1,2], Shuangyou Zhang[1], Toby Bi[1,3], George Ghalanos[1], Yaojing Zhang[1],
Haochen Yan[1,3], Arghadeep Pal[1,3], Jijun He[2], Shilong Pan[2,*], and Pascal Del'Haye[1,3,†]

[1] *Max Planck Institute for the Science of Light, 91058 Erlangen, Germany*
[2] *National Key Laboratory of Microwave Photonics, Nanjing University of Aeronautics and Astronautics, Nanjing 210016, China*
[3] *Department of Physics, Friedrich Alexander University Erlangen-Nuremberg, 91058, Germany*

**Abstract:** We demonstrate Kerr soliton frequency comb generation that is seeded by a cascaded Brillouin scattering process. In this process, a pump laser is used to generate multiple orders of Brillouin sidebands in a microresonator, which in turn generate the soliton. In such a process, even orders of Brillouin scattering sidebands are co-propagating with respect to the pump laser while odd orders of Brillouin scattering are backwards propagating. In this work we present the generation of forward propagating Kerr solitons via a forward propagating second order Brillouin scattering process in a fused silica rod resonator. Importantly, we show that the Brillouin scattering process can bridge the gap between different microresonator mode families, such that the repetition rate of the Kerr soliton is independent from the Brillouin gain frequency shift (~10 GHz in fused silica). In our work we demonstrate this by generating soliton pulse trains with a repetition rate of 107 GHz. Our work opens up a new way for using cascaded Brillouin lasing as a seed for microresonator frequency comb generation. This can be of particular interest for the realization of soliton frequency combs with low noise properties from Brillouin lasing while still having arbitrary repetition rates that are determined by the resonator size. Applications range from optical communication to LIDAR systems and photonic signal generation.

## INTRODUCTION

Dissipative Kerr solitons (DKS) that rely on the Kerr nonlinearity within high-$Q$ microresonators have received significant research interest because of their potential use in a large number of photonic applications. Soliton frequency combs have in particular garnered attention for their small foot print and small power consumption for out-of-the-lab applications [1-7]. Leveraging advanced fabrication techniques across diverse optical platforms, creating DKS with repetition rates spanning from several GHz to several THz has become possible. Attainable frequency ranges cover the microwave, millimeter-wave and terahertz domains, leading to potential applications like ultra-fast ranging [8-10], dual-comb spectroscopy [11, 12], high-speed optical communications [13], photonic signal generation [14, 15] and astro-combs [16, 17].

Significant progress has been made in DKS generation in recent years, particularly through the interaction between a stimulated Brillouin laser (SBL) [18, 19] and the nonlinear Kerr effect. This innovative approach has been demonstrated across multiple platforms, such as fiber Fabry-Pérot (FFP) cavities [20, 21], silica microdisk resonators [22] and silica wedge resonators [23]. In this concept, a pump laser generates a Brillouin sideband, which in turn generates a Kerr soliton frequency comb. This is an intriguing approach because the pump laser can remain on the thermally stable blue-detuned side of a microresonator mode. An alternative approach for stable soliton generation by balancing thermal effects has been shown by employing dual pumping using an auxiliary laser [24, 25]. This configuration achieves self-stabilization for thermal compensation and ensures the robust generation of solitons [26]. Dual pumping has enabled comb generation with improved phase noise, comb linewidth, and timing jitter [21] compared to earlier DKS schemes.

First order Brillouin scattering and thus also the generated soliton combs are backwards propagating with respect to the pump laser. So far, there is no report of SBL-Kerr soliton generation based on higher-order stimulated Brillouin scattering, which can be co-propagating with the pump light. Here, we introduce an implementation of dissipative Kerr solitons that are generated by a cascaded process of forward-propagating stimulated Brillouin scattering (SBS) within a fused silica rod [27, 28]. The experiments are done in resonators with a free spectral range (FSR) of 107 GHz, which is not linked to the SBS frequency shift. Notably, our results establish the feasibility of achieving SBS-cascaded Kerr solitons with arbitrary repetition rates conducted by exciting different-order Brillouin laser sidebands. These findings could be interesting for optical communication, ranging, and photonic microwave generation in integrated photonic platforms.

## EXPERIMENT AND ANALYSIS

Figure 1(a) shows the theoretical concept of the forward SBS cascaded comb generation. If the Brillouin gain overlaps with a resonator mode and surpasses the cavity losses, it undergoes strong amplification and becomes the dominant lasing mode. In our case, both the 1$^{st}$ and 2$^{nd}$ order SBS gain regions overlap with cavity modes, giving rise to 2$^{nd}$ order cascaded SBS. The generated SBS sidebands can give rise to four-wave mixing (FWM) induced optical frequency combs. By controlling the pump laser detuning, the SBS sideband starts to generate a low noise dissipative Kerr soliton.

To realize the proposed method, we set up the experiment shown in Fig. 1(b). An external tunable laser with power and polarization control is used as pump source. This laser is directed towards a 0.3-mm-radius silica microrod cavity [27, 28] with many higher order mode families. Many of the modes in the different mode families have a frequency offset that matches the SBS gain offset. The oscilloscope (OSC) captures the signals of both the forward and backward propagation from the microrod cavity with two photodetectors (PDs). In addition, these signals can be recorded using an optical spectrum analyzer (OSA).

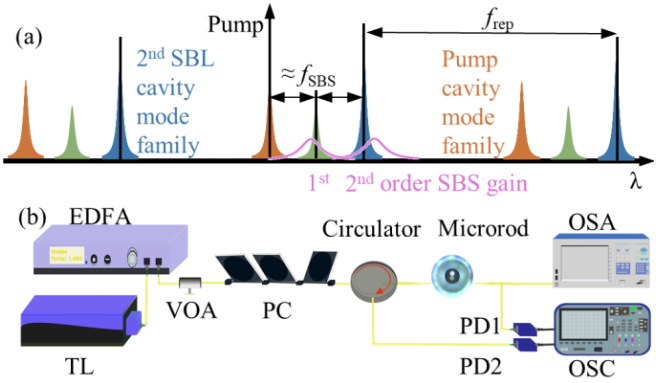

Fig. 1. (a) Schematic diagram of the forward SBS cascaded comb generation. The orange resonances represent the pump cavity mode family, the green resonances correspond to the mode family for the 1st order SBS sidebands, and the cyan resonances show the mode family that overlaps with the 2nd order SBS sideband. The pink solid curves represent the 1st order and 2nd order SBL gain regions excited by the pump laser. $f_{SBS}$: Brillouin frequency shift. $f_{rep}$: the repetition rate of the comb. (b) Experimental setup for the forward SBS cascaded soliton generation. TL: tunable laser; EDFA: erbium-doped fiber amplifier; VOA: variable optical attenuator; PC: polarization controller; PD: photodiode; OSA: optical spectrum analyzer; OSC: oscilloscope.

As shown in Fig. 2(a), we observe forward-propagating soliton states. The blue solid line is a fitted sech² envelope, which closely matches the measured spectrum. Zooming into the spectrum, as depicted in Fig. 2(b) the measured FSR is around 0.87 nm (107 GHz). The spacing between the pump cavity mode and the neighboring cavity mode for the 1st SBS sideband is 0.09 nm. Note that the first order SBS sideband is backwards propagating in the microresonators. Thus, the measured signal in forward direction in Fig 2(b) only shows a weak signal for the first order SBS sideband, which originates from backscattering within the resonator. The spacing between the pump cavity mode and the 2nd SBL cavity mode utilized for generating the SBL-Kerr soliton is 0.18 nm. This mode spacing matches the frequency shift for the 2nd order stimulated Brillouin sideband of 22.5 GHz. It should be noted that the signal measured on the by OSA has been attenuated by a 10-dB optical attenuator and optical couplers.

Figure 2(c) and 2(d) show forward transmission spectra of the microrod cavity at low and high optical pump power. The inset of Fig. 2(d) provides a close-up of the transmission of the cavity mode that overlaps with the 2nd order SBL gain. The different mode families for the pump laser, 1st and 2nd order Brillouin sidebands are marked with colored circles. Moreover, the measured linewidths of the pump mode, the mode of the 1st SBS sideband, and the mode of the 2nd SBS sideband are 13.3 MHz, 3.99 MHz, and 1.6 MHz (see Fig. 2(e)), whereas their corresponding loaded quality (Q) factors are $0.144 \times 10^8$, $0.48 \times 10^8$ and $1.2 \times 10^8$, respectively.

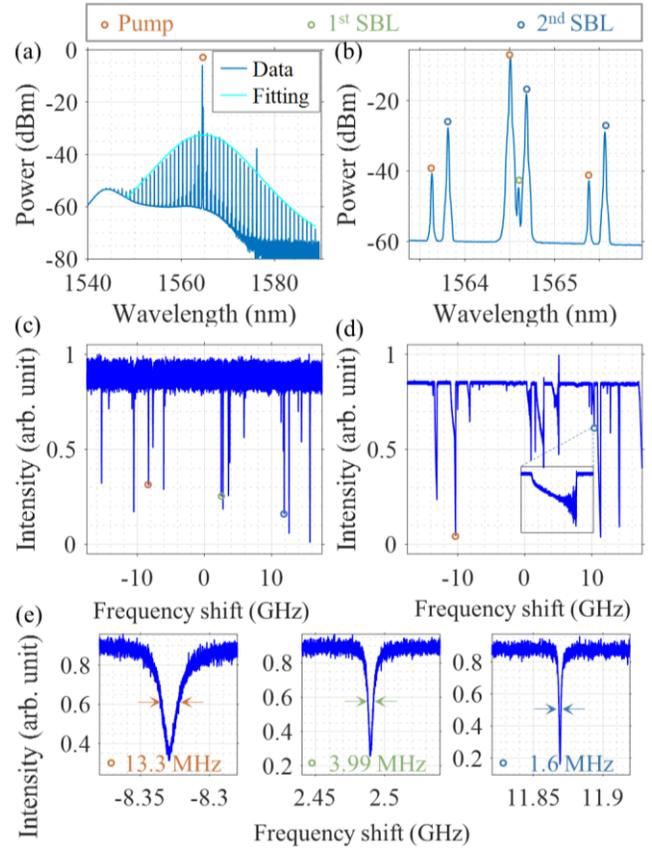

Fig. 2. (a) Measured spectrum of a forward propagating single soliton. The cyan-colored solid line is a sech² fit. (b) Close-up of the spectrum of (a). The different mode families for the pump laser, 1st and 2nd order Brillouin sidebands are marked with colored circles, according to the legend on top. (c) Forward transmission spectrum of the microrod cavity at low optical pump power. (d) Forward transmission spectrum at high optical pump power. (e) Linewidth of the pump mode (13.3 MHz), the 1st SBL cavity mode (3.99 MHz) and the 2nd SBL cavity mode (1.6 MHz).

In order to study the SBL-Kerr interactions, we scan the tunable pump laser across the pump mode with a power of 15 dBm. The corresponding transmission and reflection curves are shown in Fig. 3. In contrast to a uniform conventional thermal triangle, we can divide the transmission/reflection curves into three parts.

In region I, the intracavity buildup is less than the threshold power for the SBS and parametric oscillations. Consequently, only the pump laser without sidebands is measured on the OSA in this region. When red-detuning the laser frequency, the overlap between the SBL gain region and the SBS cavity mode increases. Once this overlap is sufficiently strong, the backward propagating 1st order SBS sideband is excited together with a forward propagating 2nd order SBS sideband, which can be seen as region II in Fig. 3. Initially, the backward intensity substantially increases due to the 1st order SBS process. When passing through region II, the overlap of the 2nd order SBS gain region and its corresponding cavity mode increases, which leads to a reduction in backwards propagating light. Consequently, in region II we can observe simultaneously the pump laser, the 1st order Brillouin lasing and the 2nd order Brillouin lasing on the optical spectrum analyzer.

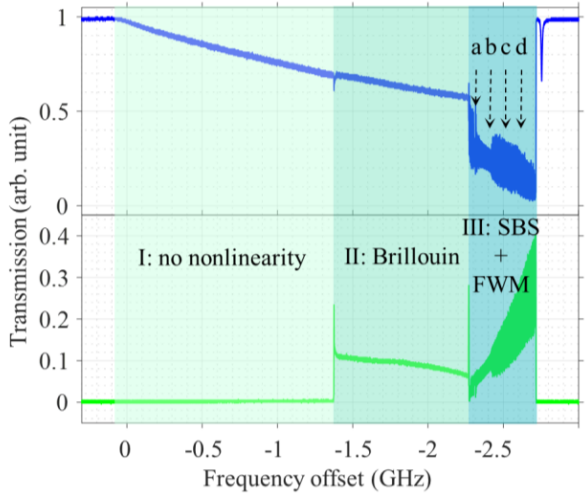

Fig. 3. Transmission (blue) and reflection (green) of the microresonator mode when scanning the pump laser from high to low frequencies. We observe three distinct regions in the thermally broadened resonance (I, II, III). Four different comb states can be accessed and are marked with dashed arrows: a (modulational instability comb), b (single soliton), c (multiple-solitons), d (perfect soliton crystal).

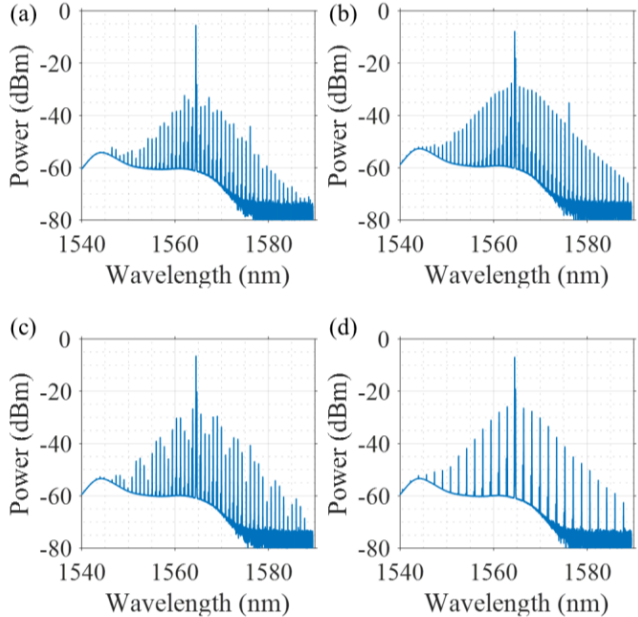

Fig. 4. The optical spectra of (a) modulational instability comb, (b) single soliton, (c) multiple-soliton, (d) perfect 2-FSR soliton crystal.

Since the measured loaded Q factor of the $2^{nd}$ order SBS cavity mode ($1.2 \times 10^8$) is much higher than that of the pump mode ($0.144 \times 10^8$), the effective loaded Q factor in the forward direction becomes better when more intracavity power is transferred to the $2^{nd}$ order SBS cavity mode in region II.

When we further red-detune the laser frequency, we enter region III. Here, parametric oscillations induced directly from the $2^{nd}$ order Brillouin lasing appear and create multiple optical sidebands. In addition, the pump laser itself generates optical sidebands through four-wave mixing. This can be seen as a chaotic modulational instability comb in Fig. 4(a). After traversing this chaotic region, we reach the soliton region, starting with a single soliton shown in Fig. 4(b), a multi-soliton state shown in Fig. 4(c), and eventually a perfect soliton crystal shown in Fig. 4(d). These soliton combs are centered around the second order SBS sideband. The faintly visible Turing pattern comb in the background of the OSA spectra is directly generated by the pump light and is spectrally offset from the soliton combs. The soliton states remain stable for several hours during our experiments. Since the exact soliton number for the bright soliton is not predetermined, a multi-soliton is typically observed upon entering the soliton region. However, we can deterministically reduce the number of solitons by using the backwards-detuning method described in Ref. [29] to obtain a single soliton state. Note that in previous work, the presence of cascaded Brillouin scattering prevented the generation of dissipative Kerr solitons [20, 30]. Here we show, that both processes can coexist within a microresonator.

## DISCUSSION AND CONCLUSION

In summary, we demonstrate a novel mechanism to generate forward propagating solitons that are induced by cascaded Brillouin scattering. Our work presents stable soliton states with a repetition rate of 107 GHz using a fused silica microrod resonator. Leveraging the abundance of different mode families within the microrod cavity, we are able to overlap the SBS gain regions with cavity modes to support the cascaded soliton generation process. Moreover, our findings introduce a novel approach for generating soliton combs in cavities with a large number of co-existing mode families. Unlike the previous SBL-Kerr soliton generation method [20-23], the soliton generation happens in a different mode family, which leads to a repetition rate that is independent of the Brillouin frequency shift. Thus, this method has the potential to generate SBL induced soliton combs at arbitrary repetition rates by using microresonators with different diameters. This approach allows us to achieve low-noise solitons via low-noise Brillouin lasing, which is essential for photonic signal generation [31] and optical atomic clock devices [32, 33]. Moreover, SBL induced soliton combs can have applications in optical communication, ranging, and photonic microwave generation.

**Acknowledgments:** The project is funded by European Union's H2020 ERC Starting Grant "CounterLight" (756966); H2020 Marie Sklodowska-Curie COFUND "Multiply" (713694); Marie Curie Innovative Training Network "Microcombs" (812818); China Scholarship Council (202106830063); National Natural Science Foundation of China (62205145); National Key Research and Development Program of China (2022YFB2802700; Natural Science Foundation of Jiangsu Province (BK20220887)

**Author contributions:** H. Z. fabricated the cavity and performed the experiments. H. Z. and S. Z. designed the experiments and analyzed the data. T. B. built the processing platform for microrod. G. G. and A.P. built the setup for fabricating tapered fibers. H. Z. wrote the manuscript, with input from the other authors. P. D. supervised the project.

† pascal.delhaye@mpl.mpg.de
* pans@nuaa.edu.cn

1. T. J. Kippenberg, A. L. Gaeta, M. Lipson, and M. L. Gorodetsky, "Dissipative Kerr solitons in optical microresonators," Science **361**, eaan8083 (2018).
2. A. L. Gaeta, M. Lipson, and T. J. Kippenberg, "Photonic-chip-based frequency combs," Nature Photonics **13**, 158-169 (2019).
3. Y. S. Jang, H. Liu, J. Yang, M. Yu, D. L. Kwong, and C. W. Wong, "Nanometric Precision Distance Metrology via Hybrid Spectrally Resolved and Homodyne Interferometry in a Single Soliton Frequency Microcomb," Phys Rev Lett **126**, 023903 (2021).
4. L. Chang, S. Liu, and J. E. Bowers, "Integrated optical frequency comb technologies," Nature Photonics **16**, 95-108 (2022).
5. A. W. Bruch, X. Liu, Z. Gong, J. B. Surya, M. Li, C.-L. Zou, and H. X. Tang, "Pockels soliton microcomb," Nature Photonics (2020).
6. H. Bao, A. Cooper, M. Rowley, L. Di Lauro, J. S. Totero Gongora, S. T. Chu, B. E. Little, G.-L. Oppo, R. Morandotti, D. J. Moss, B. Wetzel, M. Peccianti, and A. Pasquazi, "Laser cavity-soliton microcombs," Nature Photonics **13**, 384-389 (2019).
7. B. Shen, L. Chang, J. Liu, H. Wang, Q. F. Yang, C. Xiang, R. N. Wang, J. He, T. Liu, W. Xie, J. Guo, D. Kinghorn, L. Wu, Q. X. Ji, T. J. Kippenberg, K. Vahala, and J. E. Bowers, "Integrated turnkey soliton microcombs," Nature **582**, 365-369 (2020).
8. M.-G. Suh, and K. J. Vahala, "Soliton microcomb range measurement," Science **359**, 884-887 (2018).
9. P. Trocha, M. Karpov, D. Ganin, M. H. P. Pfeiffer, A. Kordts, S. Wolf, J. Krockenberger, P. Marin-Palomo, C. Weimann, S. Randel, W. Freude, T. J. Kippenberg, and C. Koos, "Ultrafast optical ranging using microresonator soliton frequency combs," Science **359**, 887-891 (2018).
10. J. Riemensberger, A. Lukashchuk, M. Karpov, W. Weng, E. Lucas, J. Liu, and T. J. Kippenberg, "Massively parallel coherent laser ranging using a soliton microcomb," Nature **581**, 164-170 (2020).
11. B. Bernhardt, A. Ozawa, P. Jacquet, M. Jacquey, Y. Kobayashi, T. Udem, R. Holzwarth, G. Guelachvili, T. W. Hänsch, and N. Picqué, "Cavity-enhanced dual-comb spectroscopy," Nature Photonics **4**, 55-57 (2009).
12. B. S. Qi-Fan Yang, Heming Wang, Minh Tran, Zhewei Zhang, Ki Youl Yang, Lue Wu, Chengying Bao, John Bowers, Amnon Yariv, Kerry Vahala, "Vernier spectrometer using counterpropagating soliton microcombs," Science (2019).
13. L. Lundberg, M. Mazur, A. Mirani, B. Foo, J. Schroder, V. Torres-Company, M. Karlsson, and P. A. Andrekson, "Phase-coherent lightwave communications with frequency combs," Nat Commun **11**, 201 (2020).
14. J. Liu, E. Lucas, A. S. Raja, J. He, J. Riemensberger, R. N. Wang, M. Karpov, H. Guo, R. Bouchand, and T. J. Kippenberg, "Photonic microwave generation in the X- and K-band using integrated soliton microcombs," Nature Photonics, 486–491 (2020).
15. S. Zhang, J. M. Silver, X. Shang, L. Del Bino, N. M. Ridler, and P. Del'Haye, "Terahertz wave generation using a soliton microcomb," Opt Express **27**, 35257-35266 (2019).
16. E. Obrzud, M. Rainer, A. Harutyunyan, M. H. Anderson, J. Liu, M. Geiselmann, B. Chazelas, S. Kundermann, S. Lecomte, M. Cecconi, A. Ghedina, E. Molinari, F. Pepe, F. Wildi, F. Bouchy, T. J. Kippenberg, and T. Herr, "A microphotonic astrocomb," Nature Photonics **13**, 31-35 (2018).
17. M. G. Suh, X. Yi, Y. H. Lai, S. Leifer, I. S. Grudinin, G. Vasisht, E. C. Martin, M. P. Fitzgerald, G. Doppmann, J. Wang, D. Mawet, S. B. Papp, S. A. Diddams, C. Beichman, and K. Vahala, "Searching for Exoplanets Using a Microresonator Astrocomb," Nature Photonics **13**, 25-30 (2019).
18. M. Asano, Y. Takeuchi, S. K. Ozdemir, R. Ikuta, L. Yang, N. Imoto, and T. Yamamoto, "Stimulated Brillouin scattering and Brillouin-coupled four-wave-mixing in a silica microbottle resonator," Opt Express **24**, 12082-12092 (2016).
19. H. Wang, Y. H. Lai, Z. Yuan, M. G. Suh, and K. Vahala, "Petermann-factor sensitivity limit near an exceptional point in a Brillouin ring laser gyroscope," Nat Commun **11**, 1610 (2020).
20. K. Jia, X. Wang, D. Kwon, J. Wang, E. Tsao, H. Liu, X. Ni, J. Guo, M. Yang, X. Jiang, J. Kim, S. N. Zhu, Z. Xie, and S. W. Huang, "Photonic Flywheel in a Monolithic Fiber Resonator," Phys Rev Lett **125**, 143902 (2020).
21. M. Nie, K. Jia, Y. Xie, S. Zhu, Z. Xie, and S. W. Huang, "Synthesized spatiotemporal mode-locking and photonic flywheel in multimode mesoresonators," Nat Commun **13**, 6395 (2022).
22. Y. Bai, M. Zhang, Q. Shi, S. Ding, Y. Qin, Z. Xie, X. Jiang, and M. Xiao, "Brillouin-Kerr Soliton Frequency Combs in an Optical Microresonator," Phys Rev Lett **126**, 063901 (2021).
23. I. H. Do, D. Kim, D. Jeong, D. Suk, D. Kwon, J. Kim, J. H. Lee, and H. Lee, "Self-stabilized soliton generation in a microresonator through mode-pulled Brillouin lasing," Opt Lett **46**, 1772-1775 (2021).
24. S. Zhang, J. M. Silver, T. Bi, and P. Del'Haye, "Spectral extension and synchronization of microcombs in a single microresonator," Nat Commun **11**, 6384 (2020).
25. D. V. Strekalov, and N. Yu, "Generation of optical combs in a whispering gallery mode resonator from a bichromatic pump," Physical Review A **79** (2009).
26. S. Zhang, J. M. Silver, L. Del Bino, F. Copie, M. T. M. Woodley, G. N. Ghalanos, A. Ø. Svela, N. Moroney, and P. Del'Haye, "Sub-milliwatt-level microresonator solitons with extended access range using an auxiliary laser," Optica **6** (2019).
27. P. Del'Haye, S. A. Diddams, and S. B. Papp, "Laser-machined ultra-high-Q microrod resonators for nonlinear optics," Applied Physics Letters **102** (2013).
28. A. O. Svela, J. M. Silver, L. Del Bino, S. Zhang, M. T. M. Woodley, M. R. Vanner, and P. Del'Haye, "Coherent suppression of backscattering in optical microresonators," Light Sci Appl **9**, 204 (2020).
29. H. Guo, M. Karpov, E. Lucas, A. Kordts, M. H. P. Pfeiffer, V. Brasch, G. Lihachev, V. E. Lobanov, M. L. Gorodetsky, and T. J. Kippenberg, "Universal dynamics and deterministic switching of dissipative Kerr solitons in optical microresonators," Nature Physics **13**, 94-102 (2016).
30. D. Braje, L. Hollberg, and S. Diddams, "Brillouin-enhanced hyperparametric generation of an optical frequency comb in a monolithic highly nonlinear fiber cavity pumped by a cw laser," Phys Rev Lett **102**, 193902 (2009).
31. X. Xie, R. Bouchand, D. Nicolodi, M. Giunta, W. Hänsel, M. Lezius, A. Joshi, S. Datta, C. Alexandre, M. Lours, P.-A. Tremblin, G. Santarelli, R. Holzwarth, and Y. Le Coq, "Photonic microwave signals with zeptosecond-level absolute timing noise," Nature Photonics **11**, 44-47 (2016).
32. S. B. Papp, K. Beha, P. Del'Haye, F. Quinlan, H. Lee, K. J. Vahala, and S. A. Diddams, "Microresonator frequency comb optical clock," Optica **1**, 10-14 (2014).
33. Z. L. Newman, V. Maurice, T. Drake, J. R. Stone, T. C. Briles, D. T. Spencer, C. Fredrick, Q. Li, D. Westly, B. R. Ilic, B. Shen, M.-G. Suh, K. Y. Yang, C. Johnson, D. M. S. Johnson, L. Hollberg, K. J. Vahala, K. Srinivasan, S. A. Diddams, J. Kitching, S. B. Papp, and M. T. Hummon, "Architecture for the photonic integration of an optical atomic clock," Optica **6** (2019).